\documentclass{aa}  

\usepackage{graphicx}
\usepackage{txfonts}
\begin{document}

   \title{Evaporation ages: A new dating method for young star clusters}


   \author{V.-M. Pelkonen
          \inst{1},
          N. Miret-Roig
          \inst{2}
          \and
          P. Padoan\inst{1,3,4}
          }

   \institute{Institut de Ci\`{e}ncies del Cosmos, Universitat de Barcelona, IEEC-UB, Mart\'{i} i Franqu\'{e}s 1, E08028 Barcelona, Spain\\
              \email{veli.matti.pelkonen@icc.ub.edu}
              \and
              University of Vienna, Department of Astrophysics, Türkenschanzstraße 17, 1180 Wien, Austria
         \and
              ICREA, Pg. Llu\'{i}s Companys 23, 08010 Barcelona, Spain
         \and
         Department of Physics and Astronomy, Dartmouth College, 6127 Wilder Laboratory, Hanover, NH 03755, USA \\
             }

   \date{Received XXXXXX; accepted YYYYYY}

 
  \abstract
   {The ages of young star clusters are fundamental clocks to constrain the formation and evolution of pre-main-sequence stars and their protoplanetary disks and exoplanets. However, dating methods for very young clusters often disagree, casting doubts on the accuracy of the derived ages.}
   {We propose a new method to derive the kinematic age of star clusters based on the evaporation ages of their stars.}
   {The method was validated and calibrated using hundreds of clusters identified in a supernova-driven simulation of the interstellar medium forming stars for approximately 40\,Myr within a 250\,pc region.}
   {We demonstrate that the clusters' evaporation-age uncertainty can be as small as about 10\% for clusters with a large enough number of evaporated stars and small but with realistic observational errors. We have obtained evaporation ages for a pilot sample of ten clusters, finding a good agreement with their published isochronal ages.}
  {The evaporation ages will provide important constraints for modeling the pre-main-sequence evolution of low-mass stars, as well as allow for the star formation and gas-evaporation history of young clusters to be investigated. These ages can be more accurate than isochronal ages for very young clusters, for which observations and models are more uncertain.}

   \keywords{stars: formation -- stars: kinematics and dynamic -- open clusters and associations: general -- solar neighborhood -- open clusters and associations: individual: Ophiuchus -- open clusters and associations: individual: Upper Scorpius -- open clusters and associations: individual: $\beta$ Pictoris --open clusters and associations: individual: Tucana-Horologium -- methods: numerical
               }

   \maketitle
%

\section{Introduction} 

The precise determination of stellar ages is a fundamental tool in cosmology and astrophysics. The ages of the oldest globular clusters provide a lower limit to the age of the universe \citep[e.g.,][]{Chaboyer+96, Krauss+Chaboyer03, Jimenez+19}, while at the other extreme, the ages of the youngest stars are fundamental clocks to time the evolution of protoplanetary disks and to guide our understanding of the origin of planets \citep[e.g.,][]{Williams+Cieza11, Bell+13, Berger+23}.    

Stellar ages are derived by combining spectroscopic and photometric data with stellar evolution models \citep{Soderblom2010}, with the most used methods being the lithium-depletion boundary \citep{Basri+1996,Stauffer+1998,Barrado99,Manzi+08,Ramirez+12,Binks+14,Galindo-Guil+2022}, gyrochronology \citep{Barnes03,Barnes07,Meibom+09,Epstein14,Angus15,Angus+19,Curtis20,Lu+2021,Van-Lane+23}, asteroseismology \citep{Lebreton+Goupil14,Aerts2015,Aguirre+17,Bellinger19,Murphy21,Scutt+23}, or isochrone fitting in the color-magnitude diagram \citep[e.g.,][]{Naylor+Jeffries06,Mayne+Naylor08,Bonatto+Bica09,DaRio+10,Glatt+10,Bell+15,Ying+23}. The age accuracy is usually enhanced when such methods are used in combination with each other to date coeval stars in clusters, for example, when combining mass estimates from asteroseismology with isochrone fitting \citep[e.g.,][]{Miglio+21, Wang+23}. Because the cluster's distance and stellar membership are crucial inputs in their age estimation, improved distance and membership determinations from the \textit{Gaia} mission have resulted in the massive application of isochrone ages to large samples of young star clusters \citep{Randich+18, Bossini+2019, Cantat-Gaudin+2020, Dias+2021, Li+2022}.

An alternative approach to determining the age of unbound star clusters, independent of stellar-evolution models, is the analysis of their kinematics. The first attempts to derive kinematic expansion ages, based on the correlation of the positions and motions, date from several decades ago \citep{Blaauw1964}. A more sophisticated technique consists of tracing the positions of stars, members of a cluster, back in time using their current velocities and a Galactic potential. The dynamical traceback age is then the time corresponding to the highest stellar density. This technique requires the 3D positions and velocities of individual stars, which has been an observational limitation for years \citep{Brown+1997a, Ortega+2002, delaReza+2006, Ducourant+2014, Donaldson+2016, Miret-Roig+2018}. However, thanks to the precise astrometry of \textit{Gaia} \citep{GaiaColVallenari+2022} and extensive complementary radial velocity surveys like the Apache Point Observatory Galactic Evolution Experiment, APOGEE \citep{Majewski+2017}, it is now possible to obtain significantly more precise kinematic ages \citep{Crundall+19, Miret-Roig+2020b, Miret-Roig+2022b, Kerr+2022a, Kerr+2022b, Galli+2023, Couture+2023}.

Although kinematic ages are too uncertain for old clusters, they have become complementary to the isochrone fitting method (hereafter CMD) for ages $< 50$\,Myr. This is the age range for the clusters' sample selected in this study. Very young clusters ($< 5$\,Myr) may still be partly embedded in gas and dust from their parent cloud producing differential extinction on the cluster members, and they often display photometric variability widening the isochrone and the uncertainties on the age. CMD ages for the youngest clusters are also uncertain because the evolution of young accreting stars in the CMD is difficult to capture with standard evolutionary tracks where accretion is not appropriately modeled \citep{Froebrich+06,Baraffe+09,Hosokawa+11,Jensen+Haugboelle18}. 

A detailed comparison of modern dynamical traceback and CMD ages, including observational uncertainties, has shown that there is a systematic difference between the two, with the former shorter than the latter \citep{Miret-Roig+2023b}. That study suggests that the two methods have a different "time zero": a star cluster may be gravitationally bound before the dispersion of the parent gas cloud, so the time zero of the kinematic method that measures the expansion time would start a few million years after the time zero of the CMD method. If the timescale of cluster formation and gas dispersal were known, it would be possible to correct the dynamical traceback age to a similar time zero as the CMD age. Conversely, the age difference between the two methods may give us important clues about the cluster formation and gas dispersal processes.

This work proposes a new method of determining kinematic ages: instead of measuring the "expansion" age of the whole cluster, we evaluate the "evaporation" ages of its individual stars, and estimate the cluster age from that of the oldest evaporation ages. We subsequently show that this method results in ages on the order of the actual stellar ages, and we calibrated it using star clusters from a numerical simulation. We also applied the method to observational data, finding kinematic ages in reasonable agreement with the CMD ages. 

In the following, the term "star cluster" is used to refer to both bound and unbound groups of stars, irrespective of their stellar density or total mass. In \S\,\ref{simulation}, we briefly describe the star formation simulation, and in \S\,\ref{GMM} we characterize the clusters identified from the numerical data. The method is outlined in \S\,\ref{code} and then we explain how it was validated and calibrated in \S\,\ref{validation} using the simulated clusters. In \S\,\ref{observations} we explain how we applied the method to observations of real star clusters, and we finally summarize our conclusions in \S\,\ref{conclusions}.

\section{Simulation} \label{simulation}

The kinematic-age method presented in this work is tested with star clusters from the same supernova (SN) driven magneto-hydrodynamic (MHD) simulation as in \citet{Padoan+17SN_IV} and \citet{Padoan+20massive}. The reader is referred to those papers for details of the numerical methods. The 3D MHD equations are solved with the Ramses adaptive-mesh-refinement (AMR) code \citep{Teyssier02,Fromang+06,Teyssier07} within a cubic region of size $L_{\rm box}=250$\,pc, total mass $M_{\rm box}=1.9\times 10^6$\,M$_{\odot}$, and periodic boundary conditions. The initial conditions are taken from a SN-driven simulation that was integrated for 45\,Myr without self-gravity \citep{Padoan+16SN_I} with a mean density $n_{\rm H,0}=5$\,cm$^{-3}$ and a mean magnetic field $B_0=4.6$\,$\mu$G. The root mean square (rms) magnetic field generated by the turbulence has a value of 7.2\,$\mu$G and an average of $|\boldsymbol B|$ of 6.0\,$\mu$G, consistent with the value of $6.0 \pm 1.8$ $\mu$G derived from the "Millennium Arecibo 21-cm Absorption-Line Survey" by \citet{Heiles+Troland05}. 

The only driving force is from SN feedback, with SNe determined by the position and age of the massive sink particles formed when self-gravity is included (the SNe are randomly generated prior to star formation). When gravity is introduced, starting at $t=55.5$\,Myr from the initial conditions, the minimum cell size is $dx=0.0076$ pc, obtained through a uniform root-grid of $512^3$ cells and six AMR levels. With this setup, the simulation is run for a period of approximately 40\,Myr. To follow the collapse of prestellar cores, sink particles are created in cells where the gas density is larger than $10^6$\,cm$^{-3}$, according to several criteria designed to avoid creating spurious sink particles in regions where the gas is not collapsing \citep[see][for details]{Haugbolle+18imf}. When a sink particle of mass larger than 7.5\,M$_{\odot}$ has an age equal to the corresponding stellar lifetime for that mass \citep{Schaller+92}, a sphere of $10^{51}$\,erg of thermal energy is injected at the location of the sink particle to simulate the SN explosion \citet{Padoan+16SN_I}. 

With a total mass of $1.9\times 10^6$\,M$_{\odot}$, the mean column density of the simulation is 30\,M$_{\odot}$pc$^{-2}$, typical for spiral arms in the outer Galaxy \citep{Heyer+Terebey98}. In fact, a lower-resolution version of this simulation \citep{Padoan+16SN_I,Pan+16SN_II,Padoan+16SN_III} has been shown to produce dense clouds with properties consistent with those of real molecular clouds in the $^{12}$CO FCRAO Outer Galaxy Survey \citep{Heyer+98,Heyer+01}. These star-forming clouds are formed ab initio in the simulation, as a result of the large-scale dynamics driven by SNe. Although their disruption is also caused by SNe alone (HII regions and stellar winds are not included in the simulation yet), the clouds in the simulation have realistic lifetimes \citep{Lu+20SN}, comparable to estimates from observations of nearby galaxies \citep[e.g.,][]{Chevance+22,Lee+23phangs}

The simulation has generated $\sim$ 7\,000 stars with mass $> 1.0\,M_{\odot}$ and $\sim 600$ stars with mass $> 8\,\textup{M}_{\odot}$. The star formation is distributed over many different clouds with realistic values of the SFR, and the global SFR corresponds to a mean gas depletion time in the computational volume of almost 1 billion years, also realistic for a 250-pc scale \citep{Padoan+17SN_IV}. Young stars are found inside the densest filaments, while older ones have already left their parent clouds. Most of the stars in the simulation are formed in clusters, some of which have cleared their surrounding gas thanks to SN explosions of their most massive members.

\section{Identification of star clusters}\label{GMM}

For this work, we select six snapshots at time intervals of approximately 4.4\,Myr, with the first one at 17.2\,Myr after gravity is introduced in the simulation, and the final one at 39.2\,Myr. This selection yields a total sample of 26\,842 stars with mass $> 1.0\,\textup{M}_{\odot}$. The stars in each snapshot are assigned into clusters using their 6D phase space information, using a Gaussian Mixing Model (GMM) from the Scikit-learn software \citep{scikit-learn}. We explored models with different numbers of Gaussians (between 1 and 70) and chose the one where the Bayesian information criterion (BIC) stabilized to a minimum value. This procedure results in a total of 339 star clusters (35, 42, 63, 67, 65, and 67 in each of the snapshots in chronological order).

\begin{figure}
	\includegraphics[width=\columnwidth]{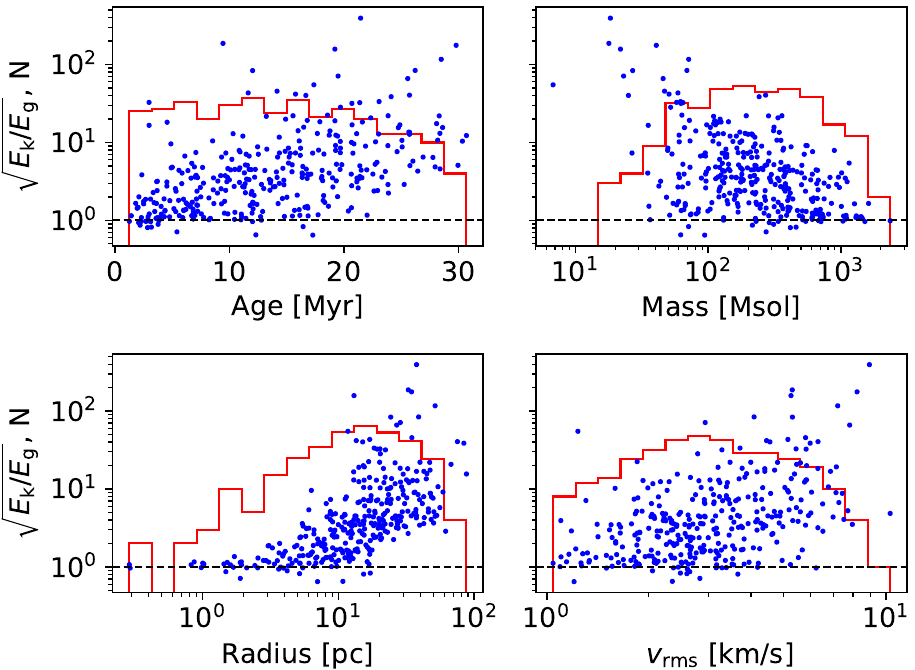}
    \caption{Histograms (in red) of the ages, masses, radius, and rms velocities and scatter plots of the square root (to use the same scale as the histograms) of the ratio of kinetic and gravitational energies versus the same quantities (blue dots) of the 339 clusters identified in the six simulation snapshots used in this work. The black dashed line marks the value of $E_{\rm k}=E_{\rm g}$, showing that 5\% of the clusters (17 of them) are gravitationally bound.}
    \label{fig:clusters}
\end{figure}

The distributions of the overall properties of our sample of star clusters are shown in the histograms in Figure\,\ref{fig:clusters}, where the cluster age is defined as the median age of the stars and the radius as the median distance of the stars to the center of mass of the cluster. The age distribution has a median value of 13\,Myr, is relatively flat in the range between approximately 1 and 20\,Myr, and decreases between 20 and 30\,Myr. The mass and radius distributions have median values of 217\,M$_{\odot}$ and 15.1\,pc, and cover the approximate ranges from 20 to 2,000\,M$_{\odot}$ and from 1 to 100\,pc, respectively. The median rms velocity of the stars in the clusters is 2.9\,km/s, ranging from 1 to 10\,km/s. We compared our numbers to the Gaia DR3 all-sky cluster catalog made by \citet{Hunt+2023}, making a cut to high-reliability, young ($<50$\,Myr), nearby ($<500$\,pc) clusters. We found that the median age was comparable to ours at 17\,Myr, while the radius was significantly smaller at 5\,pc, which is not surprising as they have a large fraction of bound clusters. On the other hand, the clusters in our observational sample presented in \S\,\ref{observations} have a median radius of 7.3\,pc but reach as high as 24\,pc, comparable to that of the simulated clusters. \citet{Castro-Ginard+2022} found a median velocity dispersion of 2.3\,km/s for their class A open clusters, which is comparable to our median velocity dispersion of 2.9\,km/s.

The scatter plots in the panels of Figure\,\ref{fig:clusters} show the square root of the ratio of kinetic and gravitational energies versus the age, mass, size, and rms velocity of the clusters. As can be seen from the $E_{\rm k}/E_{\rm g}$ ratio and the sizes, the great majority of our systems are unbound clusters that have expanded to rather large sizes. However, we also have 17 bound clusters (5\% of the total sample), and a few rather concentrated clusters with $M\sim10^3$\,M$_{\odot}$, $R\sim1$\,pc and $E_{\rm k}/E_{\rm g}\sim1$. As mentioned above, in this work we refer to both bound and unbound star groups as clusters, and our evaporation-age method works for both types of clusters.   

Because of the limited spatial resolution, the stellar initial mass function (IMF) in the simulation is incomplete below a few solar masses, and only stars $> 1.0\,\textup{M}_{\odot}$ are used in this work. The lack of low-mass stars, as well as the incomplete binary statistics, may affect the early dynamical evolution of some of the densest clusters in the simulation, while its effect on most of the diffuse clusters is probably not significant. However, because our sample contains a large number of star clusters formed and evolved ab initio from a self-consistent simulation of a very large ISM region, it represents a significant improvement over previous ad hoc models of single star clusters from pure N-body simulations without gas-dynamics, or star formation simulations of very small regions of just a few pc that were used to test kinematic-age methods \citep[e.g.,][]{Crundall+19}. Although this should be tested with future simulations including low-mass stars, we do not expect the calibration of the method to be sensitive to details of the evolution of the clusters. If anything, the inclusion of low-mass stars would make the method even more successful than concluded here, due to improved statistics. A significant fraction of the low-mass stars would undergo early evaporation, so their inclusion would increase the number of reliable evaporation ages, improving the accuracy of the method, particularly in the case of bound clusters.

\section{Evaporation-age method} \label{code}

The traditional way to determine the dynamical age of a star cluster assumes that the stars are coeval and that the cluster was never gravitationally bound, or at least started to expand very soon after its formation. The ages can then be derived by adopting a realistic Galactic potential and tracing the stellar orbits back in time to find the time of maximum stellar density \citep{delaReza+2006, Miret-Roig+2020b}, or using forward modeling of the stellar orbits to avoid propagation of observational errors \citep{Crundall+19}. Besides their uncertainties due to the observational errors, these methodologies have important limitations related to the above assumptions: i) if the cluster is initially bound for a significant time relative to its age, the kinematic age largely underestimates its real age; ii) if the cluster is still mostly gravitationally bound, the methods do not even apply (the cluster's potential is dominant over the Galactic one); iii) if the stars have a significant age dispersion relative to the age of the cluster (in the case of very young clusters) or if they become unbound at different times, the time zero of the methods is poorly defined. 

To avoid these limitations, we propose a new method of determining dynamical ages: instead of measuring the expansion age of the whole cluster, we evaluate the evaporation ages of its individual stars, and identify the cluster's age with that of the oldest evaporation ages. The evaporation age is the time since a star escaped the gravitational potential of the cluster. We define this time ignoring the cluster potential, as this requires careful modeling of the time evolution of both the gas and stars in the specific cluster, which is generally unknown, and the effect of the cluster potential is accounted for statistically when the method is calibrated with the simulated clusters. Younger evaporation ages may also correspond to the actual stellar ages if those stars were formed at a later time, or may represent a lower limit to the stellar ages if the stars had remained gravitationally bound to the system for a longer time and evaporated only more recently. For these reasons, we only use the oldest evaporation ages to establish the cluster age. However, the complete age and spatial distributions of the evaporated stars contain valuable information about the formation and dispersion process of a star cluster. 

This method has a time zero comparable to that of the CMD method if the first evaporated stars are gravitationally bound to the cluster for a short time relative to their age, and if the observations can identify them, despite their large position and velocity dispersions. The time zero of the evaporation ages is independent of the duration of the bound phase of the cluster (even bound clusters undergo evaporation) and of its age dispersion. Therefore, this method can be applied to bound clusters as long as enough evaporated stars can be identified observationally. The method becomes increasingly uncertain for clusters of increasing age or increasing escape velocity, because of the increasing size of the spatial and velocity windows that must be explored to identify the oldest evaporated stars, and because the past trajectories become more uncertain.

\begin{figure}
	\includegraphics[width=\columnwidth]{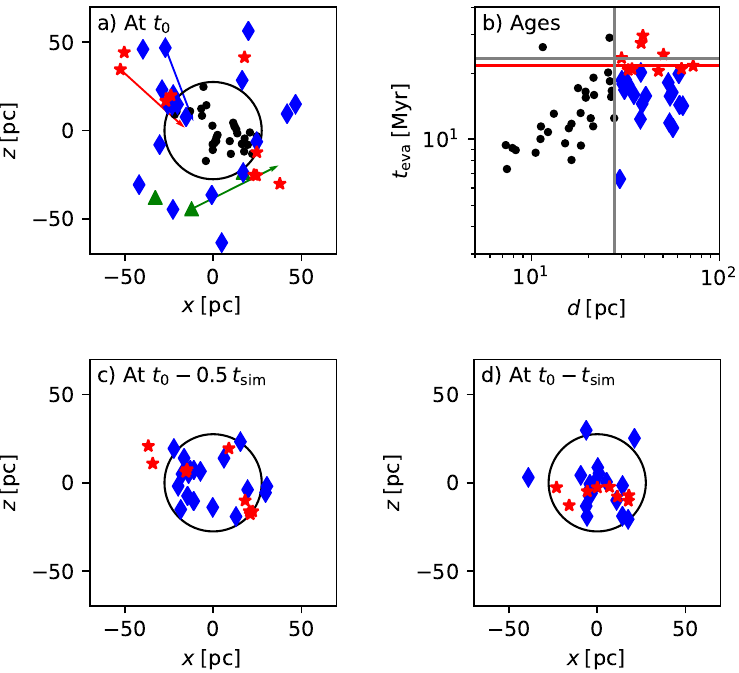}
    \caption{Example of the application of the evaporation-age method to a simulated cluster. a) Projected $xz$ positions of the cluster members in the present ($t_0$). Green triangles are initial cluster members rejected because they do not trace back into the cluster core (black circle). Black dots are cluster members in the cluster core in the present. Blue diamonds and red stars are evaporated cluster members. They are currently outside the cluster core but were inside in the past. Vectors show the 2D projected velocities back in time. b) The diagram illustrating the evaporation age determination. The x-axis is the current distance from the cluster center, $d$, with the gray vertical line showing the current core radius of the cluster, and the y-axis is the evaporation age, $t_{\rm eva}$, for each star. Red stars are evaporated stars with $t_{\rm eva}$ higher than the 70th percentile of the $t_{\rm eva}$ distribution, and the red line is the median $t_{\rm eva}$ for these stars. Blue diamonds are evaporated stars with $t_{\rm eva}$ lower than the 70th percentile and are not used for the evaporation age determination. The gray horizontal line is the simulation age of the cluster, $t_{\rm sim}$. c) Projected $XZ$ positions of evaporated stars at half $t_{\rm sim}$. d) Positions of evaporated stars at the beginning of the cluster. Only 50\% of all the cluster stars had been born by this time.}
    \label{fig:example}
\end{figure}

The specific implementation of our evaporation-age method can be summarized by the following steps that are applied to each of the individual clusters identified by the GMM in the simulation snapshots. In the case of the clusters from the simulation, we assume constant velocities equal to the current values because the simulation does not include a Galactic potential. In general, when applied to real clusters, the orbits and the cluster center should be traced back in time accounting for the Galactic potential (see \S\,\ref{observations}).  
\begin{enumerate}
    \item {\bf Rejection of divergent stars}: We determine the position and velocity of a cluster's center by taking the mean position and the mean velocity of all the stars in the cluster. The core radius, $R_{\rm 50}$, is defined as the median distance of the stars from the center. We then trace each star back in time, until it reaches its closest 3D point to the center. As we do not know apriori the age of a real cluster, we go as far back in time as necessary. If both the star and this closest point are outside of the radius, we reject the star for the rest of the analysis. 
    \item {\bf Subclustering}: It is not easy to find a single, optimal number of components for the GMM. The GMM sometimes fails to separate all the clusters, in particular younger clusters that are close together. In order to break these "joined clusters" apart to their actual individual clusters, we perform additional subclustering. We identify dense concentrations of stars, with five stars or more and a maximum mutual distance of 1 pc, as potential subcluster centers (see Appendix\,\ref{AppA} for more details). The cluster's center from the previous step is always included as the first subcluster center, too. This way, if there are no additional subclusters, the original cluster is treated as a single subcluster. We assign each star in the original cluster to the new potential subcluster it passes closest to. For each star member of the new subclusters, we calculate the closest distance to the subcluster center, the star evaporation age, defined as the time elapsed since the star was closest to the subcluster center, and the current distance to the center. With this step, a single GMM cluster may break into several subclusters, if a sufficiently high number of evaporated stars are found for more than one subcluster (see below). 
    \item {\bf Evaporation age}: For each subcluster, we select all the stars of the subcluster that trace back inside its core radius and are currently located outside; we refer to the number of these evaporated stars as $n_{\rm eva}$.
    We compute the 70th percentile of the star evaporation ages, $t_{\rm 70}$, and remove any evaporated stars with evaporation age larger than $1.4 \times t_{\rm 70} + 2 \; \rm Myr$, to reject potential outliers. We iterate until no more stars are removed and then select the stars with evaporation ages larger than the 70th percentile. We compute the evaporation age of the subcluster, $t_{\rm eva}$, as the median of the evaporation ages of this selection of stars, and record the number of stars used to obtain this value, $n_{\rm eva,70}$. We only keep the subclusters with $n_{\rm eva,70} > 1$.
\end{enumerate}

The steps of the evaporation-age method are illustrated in Figure\,\ref{fig:example}, with one of the simulated clusters as an example. The top-left panel shows the current positions of the stars, where we have identified a few divergent stars (green triangles) that do not trace back (green vector) to the core radius. The other stars (blue diamonds and red stars, $n_{\rm eva}$) trace back to the core radius (blue and red vectors) or are already inside (black points). In this particular example, we did not find subclusters, so we can directly plot the evaporation ages of individual stars as a function of the current distance, $d$, to the center, on the top-right panel. We used the stars outside the core radius to obtain the complete evaporation-age distribution, but only the stars with an evaporation age equal to or larger than the 70th percentile (red stars, $n_{\rm eva,70}$) are used to obtain the final evaporation-age of the cluster, $t_{\rm eva}$. In this example, we did not find any outliers with the $t_{\rm 70}$ criterion mentioned above, and the evaporation age that we determine (red line) is slightly underestimated compared to the simulation age (gray line). The two bottom panels show the positions of the stars that are currently outside the cluster's core radius when traced back in time to half of $t_{\rm sim}$ (bottom-left) and then by a full $t_{\rm sim}$ (bottom-right). At $t_{\rm sim}$, only half of all the stars were born and they may have been gravitationally bound. Thus, some stars overshoot the radius, when traced back to a time before their evaporation time.

\section{Validation and calibration of the method} \label{validation}

\subsection{Without observational errors}  \label{wo_errors}

We define the evaporation age of a cluster, $t_{\rm eva}$, as the median of the oldest evaporation ages of individual stars above the 70th percentile. Because different percentile values would give different ages, we need to calibrate the derived $t_{\rm eva}$ for the specific percentile value we have chosen. We calibrate and evaluate the accuracy of $t_{\rm eva}$ by comparison with the true age of the clusters in the simulation, $t_{\rm sim}$, which we define as the median age of all the stars in the cluster. We first perform this analysis by applying the method without adding position or velocity errors to the stars in the simulation. To ensure the quality of the evaporation ages, we consider only the clusters with $t_{\rm eva} > 2 \; {\rm Myr}$, as our tests showed that the accuracy decreases significantly with younger clusters. 

\begin{figure*}
	\includegraphics[width=2.0\columnwidth]{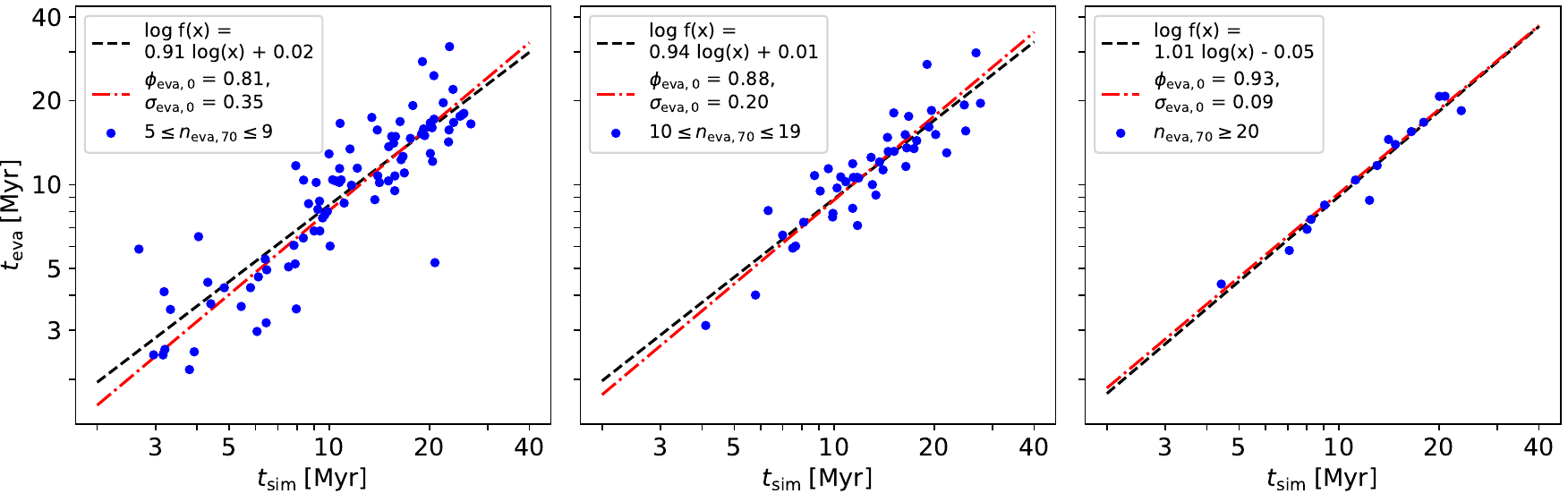}
    \caption{Evaporation age, $t_{\rm eva}$, versus simulation age, $t_{\rm sim}$, for the clusters in the simulation within three different bins of the number of stars, $n_{\rm eva,70}$, from which the evaporation age was calculated. The left panel has 86 clusters, the middle one has 45 clusters, and the right one has 15 clusters. The dashdot red line corresponds to $\phi_{\rm eva,0}\,t_{\rm sim}$, where $\phi_{\rm eva,0}$ is the median value of the age ratio $t_{\rm eva}/t_{\rm sim}$ (Eq.\,\ref{eq:phi}) in each bin. The scatter, $\sigma_{\rm eva,0}$, is the standard deviation of the ratio of the corrected $t_{\rm eva}^*$ (see Eq.\,\ref{eq:teva}) of the individual clusters with respect to their $t_{\rm sim}$, as explained in Eq.\,\ref{eq:sigma}. The black dashed line is a least-squares fit, $f(x)$, showing that the slope is very close to unity and thus $\phi_{\rm eva,0}$ does not depend on the age of the cluster.}
    \label{fig:ages}
\end{figure*}

The values of $t_{\rm eva}$ and $t_{\rm sim}$ for all the clusters in the six snapshots of the simulation are plotted in Figure\,\ref{fig:ages}, with the three panels showing clusters within three $n_{\rm eva,70}$ bins: 5-9, 10-19, and 20 or more, from left to right. The Figure shows a strong correlation between $t_{\rm eva}$ and $t_{\rm sim}$, increasing with increasing values of $n_{\rm eva,70}$. In addition, the least-squares fit (black dashed lines) has a slope very close to unity, so we can calibrate the method by dividing the derived $t_{\rm eva}$ by a scaling factor, $\phi_{\rm eva}$, independent of cluster age, which we compute as the median of all the age ratios, \begin{equation} \label{eq:phi}
\phi_{\rm eva}={{\rm med}\{t_{\rm eva}/t_{\rm sim}\}},
\end{equation}
shown by the red dashed-dotted lines in Figure\,\ref{fig:ages}. We will derive the full dependence of $\phi_{\rm eva}$ on both $n_{\rm eva,70}$ and the size of the position and velocity errors in \S\,\ref{w_errors}. For the three $n_{\rm eva,70}$ bins shown in Figure\,\ref{fig:ages}, the calibration factor without observational noise, $\phi_{\rm eva,0}$, varies from 0.81 to 0.93. 

With the established calibration, the corrected evaporation age of a cluster, $t_{{\rm eva}}^*$, is defined as
\begin{equation} \label{eq:teva}
t_{{\rm eva}}^* = t_{{\rm eva}}/\phi_{\rm eva},
\end{equation}
and the statistical uncertainty, $\sigma_{\rm eva}$, is estimated as the standard deviation of the ratio of the individual corrected ages divided by their simulation ages, $t_{{\rm eva}}^*/t_{{\rm sim}}$, a measure of the scatter in Figure\,\ref{fig:ages}, 
\begin{equation}  \label{eq:sigma}
\sigma_{\rm eva} = {{\rm std}\{t_{{\rm eva}}^*/t_{{\rm sim}}\}}.
\end{equation}
As seen in Fig.\,\ref{fig:ages}, this uncertainty is actually symmetrical in logarithmic space rather than in linear space. Thus, it should be interpreted as a factor: if $\sigma_{\rm eva}=0.3$, the upper $1\sigma$ limit should be $1.3 \times t^*_{\rm eva}$, while the lower limit should be $t^*_{\rm eva}/1.3$.

As the parameter $\phi_{\rm eva}$, $\sigma_{\rm eva}$ is also approximately independent of cluster age for $t_{\rm eva}>2$\,Myr, and it only depends on the number of evaporated stars ($n_{\rm eva,70}$) and on the size of the position and velocity errors (see  \S\,\ref{w_errors}). For the three $n_{\rm eva,70}$ bins shown in Figure\,\ref{fig:ages}, $\sigma_{\rm eva}$ is quite sensitive to $n_{\rm eva,70}$, decreasing from 35\% to 9\%. Thus, with sufficiently small observational errors, the evaporation-age method could in principle yield ages with a statistical error approaching 9\%. Because $n_{\rm eva,70}$ is defined by the 70th percentile, achieving a 9\% error or lower requires the identification of at least 67 stars outside of the cluster's core radius with measured evaporation ages (stars tracing back to the cluster's core), so a minimum of 134 stars in total (assuming no star is rejected). Figure\,\ref{fig:ages} shows that only a few of the clusters in our sample have $n_{\rm eva,70}$ values large enough to achieve this high accuracy. However, if we had all the lower-mass stars in the simulation, our total number of stars would increase by $\sim 10$, yielding a much larger number of clusters in the right panel of Figure\,\ref{fig:ages}.

\begin{figure}
	\includegraphics[width=1.0\columnwidth]{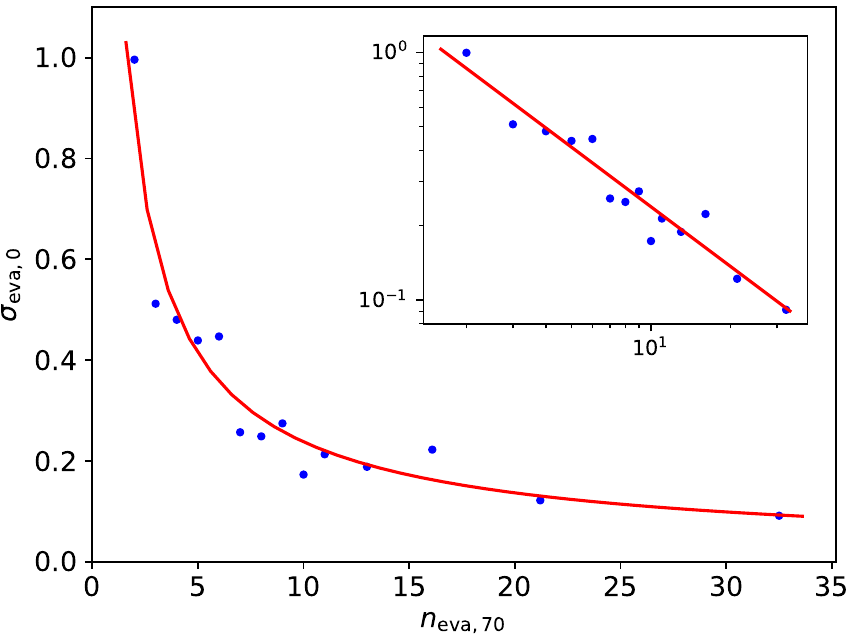}
    \caption{Scatter $\sigma_{\rm eva,0}$ as a function of $n_{\rm eva,70}$. The clusters are binned according to their $n_{\rm eva,70}$ so that each bin has ten or more clusters, with the exception of the last bin that has 8 clusters. The center of each bin is the mean of $n_{\rm eva,70}$ of the clusters in that bin. $\phi_{\rm eva}$ and $\sigma_{\rm eva}$ are then derived for each bin using Eq.\,\ref{eq:phi} and Eq.\,\ref{eq:sigma}. The least-squares fit to the [bin center, $\sigma_{\rm eva}$] pairs (blue points) is done in log-log space (see the inset panel), weighing by the square root of the number of clusters in each bin, which results in a fit $f(x) = 1.5 \; x^{-0.8}$ in linear space (red line).}
    \label{fig:sigma}
\end{figure}

Figure\,\ref{fig:sigma} shows $\sigma_{\rm eva,0}$ ($\sigma_{\rm eva}$ without observational errors) as a function of $n_{\rm eva,70}$, where the clusters are binned according to their $n_{\rm eva,70}$ so that each bin has ten or more clusters, with the exception of the last bin that has 8 clusters. We do a least-squares fit in log-log space, which results in the fitting formula:
\begin{equation} {\label{eq:sigma0}}
 \sigma_{\rm eva,0}(n_{\rm eva,70}) = 1.5 \; n_{\rm eva,70}^{-0.8},    
\end{equation}
in linear space. This formula expresses the statistical error intrinsic to the method, that is in the absence of observational errors.

\subsection{With position and velocity errors}  \label{w_errors}

In this section, we investigate the dependence of $\phi_{\rm eva}$ and $\sigma_{\rm eva}$ on the observational errors in the positions and velocities. Assuming Gaussian error distributions, we added random errors to the current positions and velocities of the stars in the simulation, with six different $1\sigma$ values in both position and velocity, hence a 6x6 grid of $1\sigma$ error pairs (in the ranges 0.15--4.80\,pc in positions and 0.15--4.80\,km/s in velocities). These values represent the present errors achievable with \textit{Gaia} only (astrometry and radial velocities) for nearby stars ($<500$\,pc) and also the \textit{Gaia} astrometry complemented with ground-based high-resolution spectroscopy. For example, in Upper Scorpius, the tangential errors are around 0.2 km/s, and with ground-based observations plus strict quality criteria (like the ones applied in our samples), it is possible to get radial velocity uncertainties of $\lesssim1$ km/s. At the same time, the uncertainties of Gaia's radial velocities are a few kilometers per second. We assumed that these error values, $\sigma_{p}$ and $\sigma_{v}$, were the same in the three directions, $\sigma_{p}=\sigma_{p,x} = \sigma_{p,y} = \sigma_{p,z}$ and $\sigma_{v} =\sigma_{v,x} = \sigma_{p,y} = \sigma_{p,z}$. In general, the observational errors in the line-of-sight direction are different (larger for nearby clusters) than in the 2D tangential plane. We have explored a case where the 1D error along the line-of-sight was three times the 1D error in another axis, and found differences in $\phi_{\rm eva}$ and $\sigma_{\rm eva}$ of only 1--10\% for the same 1D equivalent total errors, defined as:
\begin{equation}
    \sigma_{p} = \frac{1}{\sqrt{3}} \sqrt{\sigma_{p,x}^2 + \sigma_{p,y}^2 + \sigma_{p,z}^2},\\    
\end{equation}
and
\begin{equation}
    \sigma_{v} = \frac{1}{\sqrt{3}} \sqrt{\sigma_{v,x}^2 + \sigma_{v,y}^2 + \sigma_{v,z}^2}.\\    
\end{equation}
Thus, although the following calculations were carried out using equal errors in the three directions, the results hold valid for different error distributions with the same 1D equivalent total errors defined above.
We performed 1000 Monte-Carlo runs for each of the 36 error pairs, but we computed the optimal number of clusters (the GMM procedure described in \S\,\ref{GMM}) only once for each error pair to reduce the computational time.  

\begin{figure}
	\includegraphics[width=\columnwidth]{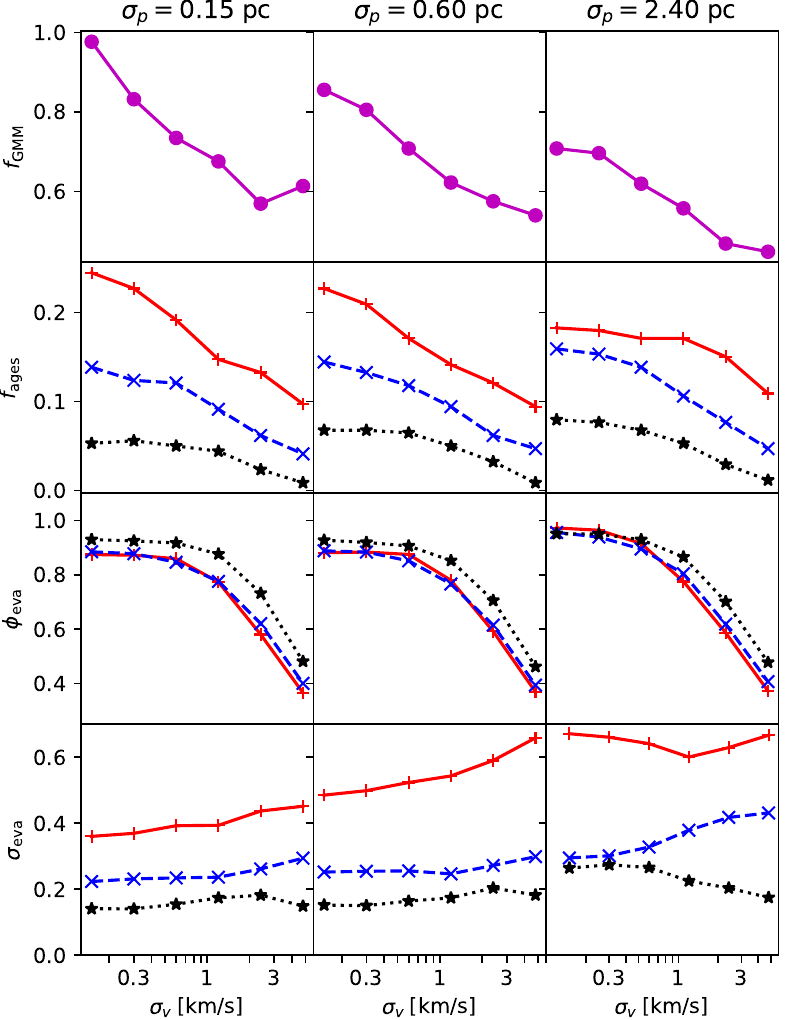}
    \caption{Impact of the observational errors. Top row: Fraction of GMM clusters, $f_{\rm GMM}$, derived from the number of clusters found by GMM in each error pair, divided by the number of GMM clusters in the no errors case (339). Second row: Fraction of clusters for which we derived ages, $f_{\rm ages}$, normalized to 339. Red, blue, and black lines are for $n_{\rm eva,70}$ bins of 5-9, 10-19, and 20+ stars, respectively. Third row: $\phi_{\rm eva}$. Bottom row: $\sigma_{\rm eva}$. Columns correspond to different 3D position errors, $\sigma_p$, and the x-axis is the 3D velocity error, $\sigma_v$. }
    \label{fig:error_plot}
\end{figure}

Figure\,\ref{fig:error_plot} shows the impact of the observational errors on the fraction of clusters that we can identify, $f_{\rm GMM}$, compared to the case with no errors), the fraction of clusters for which we can compute an evaporation age, $f_{\rm ages}$, the calibration factor, $\phi_{\rm eva}$, and statistical error, $\sigma_{\rm eva}$, of the evaporation ages. All these values are the medians of the 1000 individual realizations of each observational error pair. The fraction of identified clusters decreases significantly with increasing $\sigma_p$ and $\sigma_v$ values, going down to nearly 50\% with the largest error pair. The fraction of retained clusters, $f_{\rm ages}$, decreases with increasing $\sigma_v$, but stays constant or increases with increasing $\sigma_p$. The increase is particularly noticeable in the cases with $n_{\rm eva,70}\ge 10$ and 20. The reason is two-fold. Firstly, the larger positional errors mean that GMM sorts all the stars into fewer clusters, as seen in the top row of Figure\,\ref{fig:error_plot}, hence the individual clusters are bigger and retain more stars, so more clusters meet the minimum $n_{\rm eva,70}$ condition. Secondly, increasing random position errors cause a larger scatter of dense clumps of stars, reducing the chance that large GMM clusters get subclustered into smaller ones. 

The calibration factor, $\phi_{\rm eva}$, is not very sensitive to the value of $n_{\rm eva,70}$ or $\sigma_p$, but decreases significantly with increasing $\sigma_v$, for $\sigma_v>1$\,km/s. 
This is due to the shedding of the farthest and slowest evaporated stars: if a star is far away or has a low velocity, even a small deviation of its velocity may make it miss the cluster core and get rejected; if it is close to the cluster core or it has a high velocity the star is more likely retained. This results in a stronger reduction for the cluster's evaporation age than for the cluster's simulation age, as the latter is less sensitive to the loss of the oldest evaporated stars (they may not even be the oldest stars in the cluster). 
The value of the statistical age uncertainty, $\sigma_{\rm eva}$, increases with increasing position and velocity errors, as expected, with some exceptions in the case of $n_{\rm eva,70}>20$, due to the poor statistics (too few clusters). As in the case without errors, $\sigma_{\rm eva}$ decreases with increasing $n_{\rm eva,70}$, approaching 10\% for small enough observational errors and large enough number of stars. Histograms of $\phi_{\rm eva}$ and $\sigma_{\rm eva}$ for all 36 error pairs are given in Appendix\,\ref{AppB}, as well as the values of the medians of those histograms for five bins of $n_{\rm eva,70}$. 

\begin{figure}
 	\includegraphics[width=1.0\columnwidth]{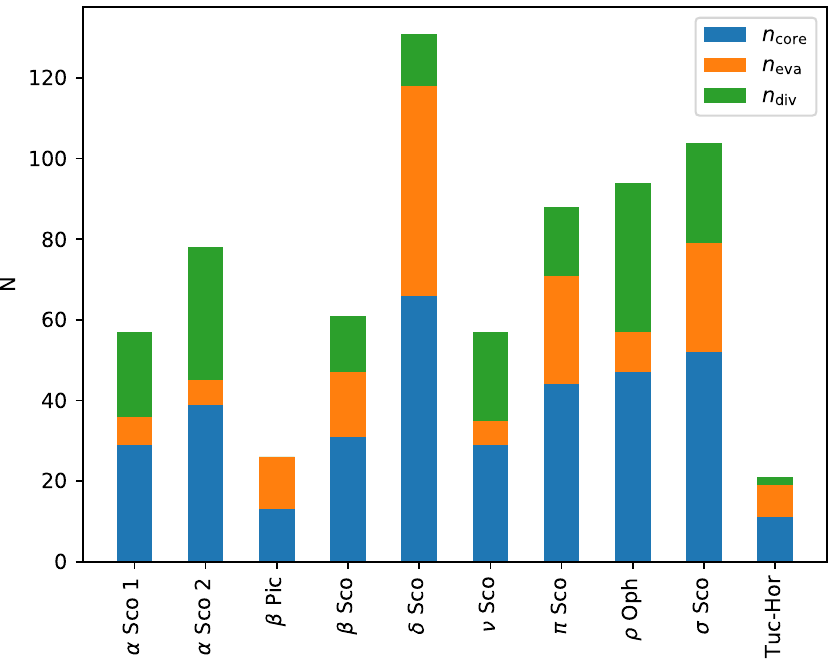}
    \caption{Number of stars in each of the ten observed clusters found within the core radius, $n_{\rm core}$ (blue), found outside of the core radius and tracing back to it, $n_{\rm eva}$ (orange), or not tracing back to it, $n_{\rm div}$ (green).
    }
    \label{fig:obs_stars}
\end{figure}

\renewcommand{\arraystretch}{1.5}
\begin{table*}
    \begin{center}
    \caption{Properties of the clusters considered in this study.}
    \label{tab:Table1}
    \centering
    \setlength{\tabcolsep}{4pt}
    \begin{tabular}{l c c c c c c c c c c c c c c }
    \hline \hline
        Cluster & $d$ & $t_{\rm CMD}$ & $t_{\rm DT}$ & $t_{\rm eva}^*$ & $\sigma_p$ & $\sigma_v$ & $R_{\rm 50}$ & $n_{\rm stars}$ & $n_{\rm div}$ & $n_{\rm eva}$  & $n_{\rm eva, 70}$ & $\phi_{\rm eva}$ & $\sigma_{\rm eva}$ \\
         & [pc] & [Myr] & [Myr] & [Myr] & [pc] & [km/s] & [pc] & & & & & & & \\
         \hline
         
$\alpha$ Sco 1 & 148 & $ 10.5^{+0.8}_{-0.4}$ & $  1.0^{+1.2}_{-1.2}$\,(b) & $ 12.1^{+13.6}_{-6.4}$ & 0.31 & 0.07 &   7.8 & 57 & 21 & 7 & 2 & 0.93 & 1.12 \\
$\alpha$ Sco 2 & 148 & $ 10.5^{+0.8}_{-0.4}$ & $  1.0^{+1.2}_{-1.2}$\,(b) & $  7.6^{+8.8}_{-4.1}$ & 0.62 & 0.10 &   6.8 & 78 & 33 & 6 & 2 & 0.89 & 1.16 \\
$\beta$ Sco & 153 & $  8.1^{+0.2}_{-0.2}$\,(a) & $  2.5^{+1.6}_{-1.6}$\,(b) & $ 10.0^{+7.4}_{-4.3}$ & 0.53 & 0.12 &   4.7 & 61 & 14 & 16 & 4 & 0.88 & 0.74 \\
$\delta$ Sco & 142 & $  7.2^{+0.3}_{-0.2}$\,(a) & $  4.6^{+1.1}_{-1.1}$\,(b) & $  7.1^{+1.7}_{-1.4}$ & 0.42 & 0.09 &   4.7 & 131 & 13 & 52 & 16 & 0.89 & 0.24 \\
$\nu$ Sco & 139 & $  5.5^{+1.3}_{-0.3}$\,(a) & $  0.3^{+0.5}_{-0.5}$\,(b) & $  4.2^{+4.8}_{-2.3}$ & 0.39 & 0.13 &   2.0 & 57 & 22 & 6 & 2 & 0.91 & 1.14 \\
$\pi$ Sco & 123 & $ 20.6^{+1.0}_{-3.2}$ & $  6.3^{+1.4}_{-1.4}$\,(b) & $ 11.1^{+4.1}_{-3.0}$ & 0.26 & 0.13 &  15.7 & 88 & 17 & 27 & 8 & 0.88 & 0.37 \\
$\sigma$ Sco & 152 & $  8.5^{+0.6}_{-0.5}$ & $  2.1^{+0.7}_{-0.7}$\,(b) & $  7.0^{+3.4}_{-2.3}$ & 0.63 & 0.14 &   9.8 & 104 & 25 & 27 & 6 & 0.88 & 0.49 \\
$\rho$ Oph & 139 & $  4.3^{+0.6}_{-0.5}$\,(a) & $  0.0^{+0.3}_{-0.3}$\,(b) & $  2.9^{+2.1}_{-1.2}$ & 0.52 & 0.10 &   3.3 & 94 & 37 & 10 & 3 & 0.88 & 0.74 \\
$\beta$ Pic &  40 & $ 20.2^{+2.2}_{-1.9}$\,(a) & $ 18.5^{+2.2}_{-2.2}$\,(c) & $ 25.8^{+15.5}_{-9.7}$ & 0.13 & 0.29 &  24.3 & 26 & 0 & 13 & 4 & 0.87 & 0.60 \\
Tuc-Hor &  47 & $ 41.8^{+3.4}_{-2.2}$\,(a) & $ 38.5^{+1.6}_{-8.0}$\,(d) & $ 53.5^{+32.9}_{-20.4}$ & 0.03 & 0.40 &  17.1 & 21 & 2 & 8 & 3 & 0.87 & 0.61 \\

         \hline
    \end{tabular}\\
    \end{center}
    \noindent \textbf{Notes.} Columns indicate (1) the cluster name, (2) the distance to the cluster, (3) the CMD age from the PARSEC isochrones used in this work, (4) the dynamical traceback age, (5) the corrected evaporation age determined in this work, (6--7) the observational error in positions and velocities, (8) the core radius, (9) the number of stars used to obtain the CMD and dynamical traceback ages and the starting sample for the evaporation ages, (10) the number of divergent stars, (11) the number of evaporated stars, (12) the number of stars used to compute the evaporation age, and (13--14) the correction factor and the uncertainty of the evaporation age (see App.\,\ref{AppB}).\\
    \noindent \textbf{References.} (a)\,\citet{Miret-Roig+2023b}; (b)\,\citet{Miret-Roig+2022b}; (c)\,\citet{Miret-Roig+2020b}; (d)\,\citet{Galli+2023}.\\
\end{table*}

\section{Results for observed clusters} \label{observations}

\subsection{Data and method}

In this section, we apply the evaporation-age method to nine young ($<50$\,Myr), nearby ($<150$\,pc) clusters, namely, $\beta$ Pictoris \citep{Miret-Roig+2020b}, Tucana-Horologium \citep{Galli+2023}, and seven clusters in Upper Scorpius and Ophiucus detailed in \citet{Miret-Roig+2022b}. We selected clusters for which we have a recent dynamical traceback age, $t_{\rm DT}$, obtained from \textit{Gaia} astrometry plus a precise selection of radial velocities (errors $\lesssim 0.5$\,km/s) from devoted ground-based observations, APOGEE, and \textit{Gaia}. To obtain the 3D Cartesian heliocentric positions and velocities for the stars in our sample, we used the \textit{Gaia} DR3 astrometry and the same radial velocities used for the dynamical traceback ages. This selection was carefully made to include only precise radial velocities ($\lesssim 1$\,km/s) and exclude potential binary stars which hinder the traceback.

A CMD age, $t_{\rm CMD}$, was also computed for these clusters, using the same stars as for the traceback analysis, and the isochrone-fitting methodology presented in \cite{Ratzenbock+2023}, using the PARSEC v1.2S models \citep{Marigo+17}. For most clusters, the values of $t_{\rm CMD}$ were presented in \citet{Miret-Roig+2023b}, while for $\alpha$\,Sco, $\pi$\,Sco and $\sigma$\,Sco they are reported here for the first time. The membership and dynamical traceback age for these three clusters are less accurate since $\alpha$\,Sco contains two subclusters identified in \citet{Ratzenbock+2022}, $\pi$\,Sco is incomplete, and $\sigma$\,Sco is more contaminated \citep{Miret-Roig+2022b}. In this work, we used the GMM code to separate $\alpha$\,Sco into two subclusters, $\alpha$\,Sco 1 and 2, by imposing a two-component solution on the original membership list, and apply the evaporation-age method to the two subclusters separately. The properties of all the clusters are summarized in Table\,\ref{tab:Table1}.

To apply the method described in \S\,\ref{code}, the orbits of the stars and the cluster center are traced back in time using a 3D Galactic potential. We tested three axisymmetric potentials, namely \textit{MWPotential14} \citep{Bovy15}, \textit{McMillan17} \citep{McMillan17}, and \textit{Irrgang13} \citep{Irrgang+13}. The differences in the evaporation ages between the Galactic potential models are very small, generally of $<1\%$, and only up to a few percent at most. The results that we report are obtained with the \textit{MWPotential14}, following the orbits back by a maximum of 50\,Myr (100\,Myr in the case of Tuc-Hor), which is enough to find the evaporation ages. The initial number of stars used to determine the evaporation ages is the same as that used for the DT and the CMD ages ($n_{\rm stars}$ in Table\,\ref{tab:Table1}), except for the subclusters of $\alpha$\,Sco, where the cluster was split in two. This sample is limited to stars with precise radial velocities, essential for the traceback analysis, but is only a lower limit to the total number of cluster members. 
Figure\,\ref{fig:obs_stars} shows the number of core, evaporated, and divergent stars for each cluster. Thanks to the careful selection of members (and 6D phase-space data) in these clusters, the number of stars rejected (divergent) in the evaporation-age method is relatively small (see Table\,\ref{tab:Table1}). The exceptions are the $\alpha$\,Sco and $\sigma$\,Sco clusters, which are likely more contaminated (see \citealt{Miret-Roig+2022b}), and $\rho$\,Oph and $\nu$\,Sco, which are the youngest groups in the sample and might still be gravitationally bound. By contrast, the more isolated clusters $\beta$\,Pic and Tuc-Hor only have two divergent stars (in Tuc-Hor); this shows the impact of the proximity of other clusters to assigning the correct membership to the stars.

\begin{figure*}
   	\includegraphics[width=2.0\columnwidth]{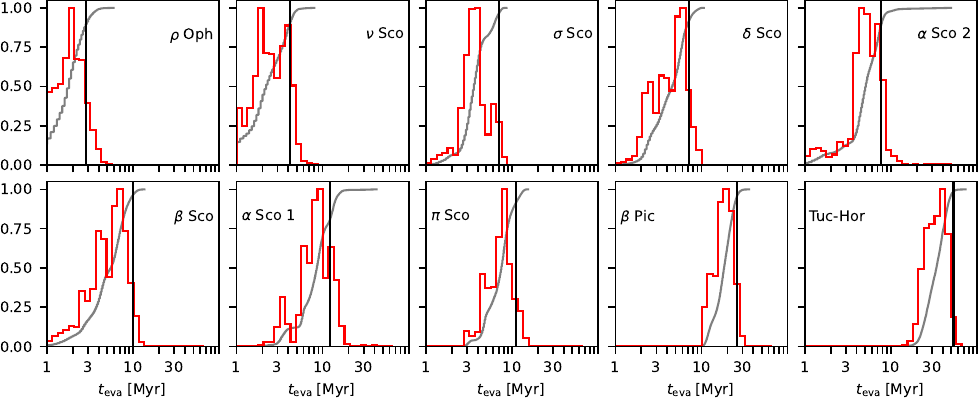}
    \caption{Histograms (peaks scaled to 1.0) of the evaporation ages of individual evaporated stars (red histograms) and the corresponding cumulative distributions (gray curves), from the 1000 Monte-Carlo error realizations and using stellar orbits computed with \textit{MWPotential14}. The derived $t^*_{\rm eva}$ is shown by the vertical black lines. 
    }
    \label{fig:obs_hist}
\end{figure*}

We derived the calibration factor, $\phi_{\rm eva}$, the corrected age, $t_{\rm eva}^*$, and its uncertainty, $\sigma_{\rm eva}$, using the 1D equivalent total observational errors and the values of $n_{\rm eva,70}$ as described in \S\,\ref{validation}. All of these values are reported in Table\,\ref{tab:Table1}. As explained in \S\,\ref{validation}, the age uncertainty is symmetrical in logarithmic space, hence used as a factor to get the error bars in linear space.

\subsection{Evaporation ages}

Figure\,\ref{fig:obs_hist} shows the distributions of the evaporation ages of individual stars (red histograms), obtained from the 1000 error realizations of each cluster. The corresponding cumulative distributions are also shown (gray curves), as well as the estimated age, $t_{\rm eva}^*$ (black vertical lines). Because the method estimates the cluster age from the oldest evaporation ages of individual stars, the peak of the distribution is expected to be at younger ages than $t_{\rm eva}^*$ (to the left of the black vertical line), as shown in Figure\,\ref{fig:obs_hist}. The difference between the time of the peak and $t_{\rm eva}^*$, typically a few Myr, is a rough estimate of the characteristic time during which the evaporated stars were bound to the cluster, which should increase over time, as more recently evaporated stars are gradually released from the cluster's potential. In addition, the width of the distribution sets an upper limit to the actual age spread of the stars (it would be equal to the age spread only if the stars were never gravitationally bound to the cluster).
Future studies should attempt a detailed analysis of the distributions of the evaporation ages of individual stars, including a comparison of the evaporation-age spread with the age spread inferred from the clusters' CMD.

\begin{figure*}
 	\includegraphics[width=2.0\columnwidth]{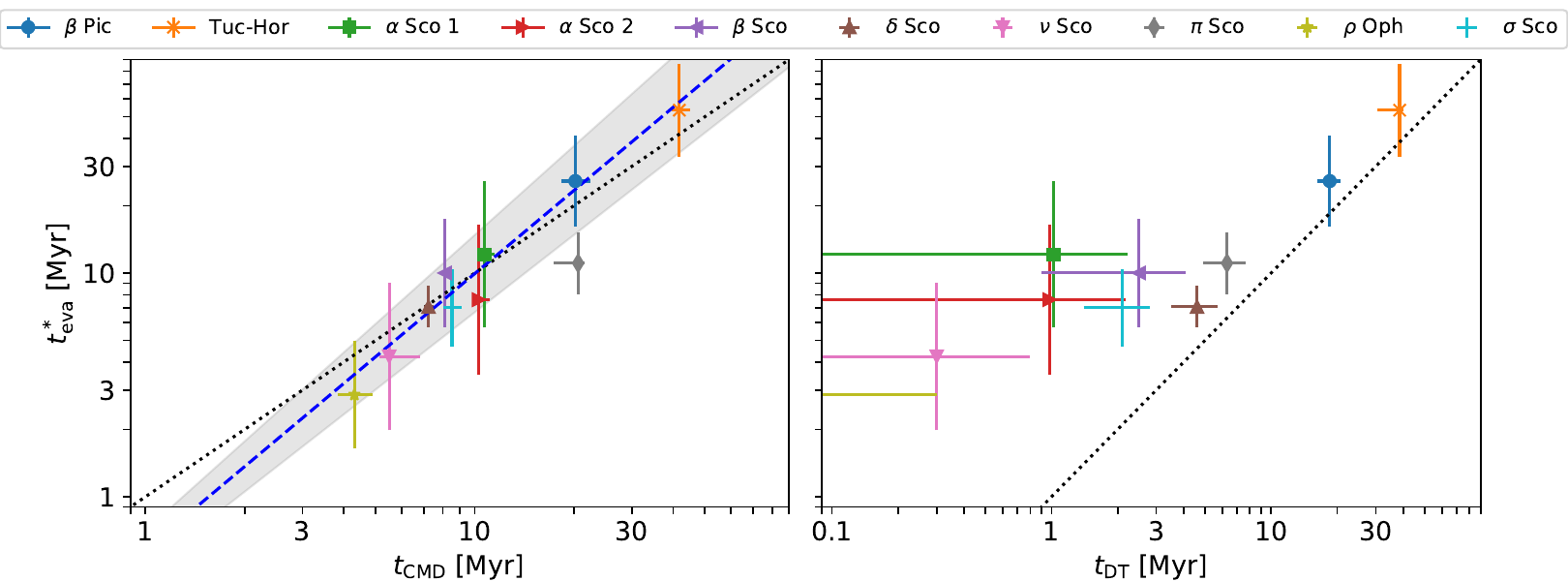}
    \caption{Comparison of the corrected evaporation ages, $t^*_{\rm eva}$, with the color-magnitude diagram ages, $t_{\rm CMD}$, and the dynamical traceback ages, $t_{\rm DT}$. The dotted black line is one-to-one line, showing that the evaporation ages are in general agreement with the CMD ages, but the DT ages are clearly offset. On the left panel, the blue dashed line is a least-squares fit to the data (except for $\pi$\,Sco) in logarithmic space, $y = 1.23^{+0.09}_{-0.09} x - 0.24^{+0.09}_{-0.09}$, with the shaded region showing the 1-sigma confidence of the fit. Apart from the youngest ages at $<3$\,Myr, the 1-to-1 line is still within the shaded region. The values used and references can be found in Table\,\ref{tab:Table1}. The $t_{\rm CMD}$ and $t_{\rm DT}$ ages of $\alpha$\,Sco\,1 and 2 have been shifted right and left by a factor of 1.02 to prevent the error bars from overlapping; the age reported in the table is in between. On the right panel, the symbol for $\rho$\,Oph is out of the plot to the left, as it has $t_{\rm DT}=0$.}
    \label{fig:obs_ages}
\end{figure*}

Figure\,\ref{fig:obs_ages} shows the comparison of the evaporation ages, $t_{\rm eva}^*$, with the CMD ages, $t_{\rm CMD}$ (left panel), and the dynamical traceback ages, $t_{\rm DT}$ (right panel). Nine out of ten clusters have an evaporation age compatible with the CMD age within the 1-$\sigma$ uncertainty. Only for $\pi$\,Sco
the evaporation age is not compatible with the CMD age. However, the census of $\pi$\,Sco is incomplete, limited to the stars inside the field of view analyzed in \citealt{Miret-Roig+2022b}. The truncation of the field of view leads to a shift of its center toward one side, reducing the stellar evaporation ages for the stars on that side and hence reducing the cluster's evaporation age. This example stresses the importance of obtaining clusters' membership lists as reliable and complete as possible.

In the left panel of Figure\,\ref{fig:obs_ages}, we have performed a least-squares fit to the data in the logarithmic scale, discarding $\pi$\,Sco due to the previously stated reasons. We find a slope of 1.2, although the 1-to-1 line is still within the 1-sigma confidence of the fit. We caution that this slope is based on mainly the two lowest and the two highest age clusters. For the two young clusters, it is possible that we miss some of the evaporated cluster members due to the crowded nature of Upper Scorpius, or perhaps for these clusters there was a longer bound period before any stars were evaporated, either of which would easily explain the shift of about 1\,Myr. Also, in all four cases, the clusters' evaporation ages are relying on just 2-4 evaporated stars above the 70th percentile. Thus, the statistics are quite uncertain.

As shown in the right panel of Figure\,\ref{fig:obs_ages}, the evaporation ages are always larger than the dynamical traceback ages. This is expected since our method is designed to find the time when the first stars evaporated. In contrast, the dynamical traceback age is designed to find the overall expansion time of the cluster, which may start well after the evaporation of the first stars. The overall expansion is better represented by the evaporation of the overall stellar population, whose peak evaporation time (the peak of the red histograms in Figure\,\ref{fig:obs_hist}) is a few million years lower than the cluster's age, $t_{\rm eva}^*$, as mentioned above. This result confirms the recent findings from the comparison between CMD and dynamical traceback ages by \citet{Miret-Roig+2023b}. 

We have found that the evaporation-age method results in ages compatible with those from isochrone fitting in the CMD. With accurate enough observations (e.g., position errors $<0.6$\,pc and velocity errors $<0.6$\,km/s) and a large enough number of stars ($n_{\rm eva,70}\ge 20$), the evaporation-age uncertainty could approach a value as low as 10\%, comparable to the typical statistical uncertainty in $t_{\rm cmd}$ from isochrone-fitting. The real uncertainty in $t_{\rm cmd}$, particularly for $t_{\rm cmd}<10$\,Myr, is significantly larger than 10\%, due to uncertainties from pre-main sequence modeling, such as the effect of starspots and magnetic fields \citep[e.g.,][]{Simon+19,Somers+20,Cao+22}, or the effect of extended and variable accretion histories \citep[e.g.,][]{Froebrich+06,Baraffe+09,Hosokawa+11,Jensen+Haugboelle18}. However, the good correlation between our evaporation ages and the CMD ages may suggest that systematic errors in the latter may not be as large as the full range of age values allowed by the largest differences between evolutionary models. As follow-up observations continue to increase the samples of accurate radial velocities in young clusters, isochrone-fitting ages should be carefully compared with evaporation ages, making sure that the same stars are used in the two methods. This comparison may shed light on the pre-main-sequence evolution (e.g., the duration of the accretion phase), on the age spread within a cluster, and on the gas-dispersal mechanism.

\section{Conclusions} \label{conclusions}

We have presented a new method to measure the age of young star clusters, based on the evaporation age of their individual stars. The method has been validated and calibrated with hundreds of star clusters with ages between 2 and 30\,Myr, taken from a star formation simulation in a region of 250\,pc, evolved for approximately 40\,Myr with self-consistent energy injection by SNe. It has also been applied to ten real clusters with previous estimates of CMD and dynamical traceback ages. Thanks to the simulated clusters, and to thousands of realizations of random observational errors, we have estimated the statistical uncertainty of the derived age, as well as a calibration factor to convert it into the median cluster age. The main conclusions of this study are listed in the following.

\begin{enumerate}

\item Using the clusters from the simulation, without adding observational errors, we find a strong correlation between the estimated age of a cluster and the median simulation age of its stars. We derive the ratio between estimated and real age, $\phi_{\rm eva,0}$, which we use to calibrate the method's age.

\item Using the simulation we also estimate the intrinsic (not due to observational errors) statistical uncertainty of the method, $\sigma_{\rm eva,0}$. We find it has no significant age dependence, and provide a fitting function for its dependence on the number of evaporated stars. We show that $\sigma_{\rm eva,0}$ can be as small as 9\% for clusters with a large number of evaporated stars.

\item By adding observational errors to the simulated clusters, we derive both $\phi_{\rm eva}$ and $\sigma_{\rm eva}$ as a function of both the number of evaporated stars and the observational errors in position and velocity, with $\sigma_{\rm eva}$ being the total uncertainty of the derived age, including both the intrinsic and observational random errors. 

\item We find that, assuming small observational errors ($\sigma_p<1$\,pc and $\sigma_v<1$\,km/s) similar to the ones achievable with \textit{Gaia} astrometry plus complementary high-resolution spectroscopy, cluster ages have a 1-$\sigma$ uncertainty $<30$\% for clusters with a minimum of $\sim 70$ stars, and approaching 10\% for clusters with $\sim 150$ stars or more, if only a small fraction of the stars are rejected. 

\item A pilot application to ten nearby clusters results in evaporation ages consistent with CMD ages, and larger than previously determined dynamical traceback ages. The uncertainties are currently significantly larger than 10\%, due to the limited number of confirmed evaporated stars in the observational samples.

\end{enumerate}

Given the large number of current and upcoming radial-velocity surveys complementing Gaia's database, evaporation ages with accuracy close to 10\% may become the norm in a few years. However, it will be necessary to explore rather wide space and velocity windows around each cluster, to capture a significant fraction of the evaporated stars, particularly the earliest ones. In practice, this will limit the method to relatively young clusters, probably with ages <50-100\,Myr, for which we can determine membership and precise stellar orbits back in time of their earliest evaporated stars. Because of theoretical uncertainties (e.g., magnetic fields and starspots) and stochasticity (e.g., extended and episodic accretion) of the pre-main sequence evolution of stars, CMD ages for very young clusters (<10-20\,Myr) are very uncertain, despite the often-quoted statistical error of $\sim$10\% from isochrone fitting. As it is completely independent of stellar evolutionary models, and particularly suited for very young clusters, the evaporation-age method will provide important constraints for the modeling of pre-main sequence evolution.

\begin{acknowledgements}
      We thank the anonymous referee for their helpful comments, leading to an improved and more clear presentation of our work.
      VMP and PP acknowledge financial support by the grant PID2020-115892GB-I00, funded by MCIN/AEI/10.13039/501100011033 and by the grant CEX2019-000918-M funded by MCIN/AEI/10.13039/501100011033. 
\end{acknowledgements}



%
%

\bibliographystyle{aa}
\bibliography{miret-roig} 
\bibliography{MC} 
\bibliography{padoan} 



\begin{appendix}

\section{Subclustering code}\label{AppA}

The subclustering code was written as a quick way to break apart mainly young, dense clusters in the simulation data that were clumped together by GMM. Thus, it is a simple code, trying to find alternative centers based on the overdensity of stars in 3D space.

\begin{enumerate}
    \item {\bf Find centers}: The first center by default is the mean location of all the stars in the cluster. After that, the code goes to each star, and finds how many stars are within 1\,pc of it. This list of stars is then sorted by the number of stars: if the number of stars is higher than five, it is a possible center. 
    \item {\bf Refining the highest overdensity center}: The code starts from the star with the highest number of neighbors, and recalculates the center coordinates based on those stars, iterating until the shift in coordinates is less than 0.1\,pc to the previous center (originally centered on a star). Then the radius is allowed to grow from 1\,pc in 0.1\,pc increments, and as long as the volume average of the stars is higher than 1.25 stars/pc$^3$, the radius continues to grow. This is to try to catch more of the stars that are on the outer edge of an overdensity. Once the criterion is no longer fulfilled, the center radius is set. Then the center position is iterated again, and then all the stars within the expanded radius are flagged as "used", so that they cannot provide new centers. We note that this radius is not the same as the core radius, $R_{\rm 50}$, used in the traceback code. 
    \item {\bf Refining the other centers}: The code goes through the next highest overdensity that remains, as in the previous step, until there are no overdensities with more than four stars within 1\,pc. 
    \item {\bf Resulting list of centers}: The list of centers and their coordinates is then passed on to the traceback code, where each star's closest pass to every center position is calculated. The star is assigned to the center that it passes closest to, as explained in the main text.
\end{enumerate}

Because the subclustering code is solely finding overdensities in 3D space, it cannot break apart two diffuse clusters. Obviously, more sophisticated clustering algorithms could be used, and in the case of individual clusters, the observers can be more careful with assigning cluster membership.

\section{Determining $\phi_{\rm eva}$ and $\sigma_{\rm eva}$ from Monte-Carlo runs}\label{AppB}

\begin{figure*}
	\includegraphics[width=2.0\columnwidth]{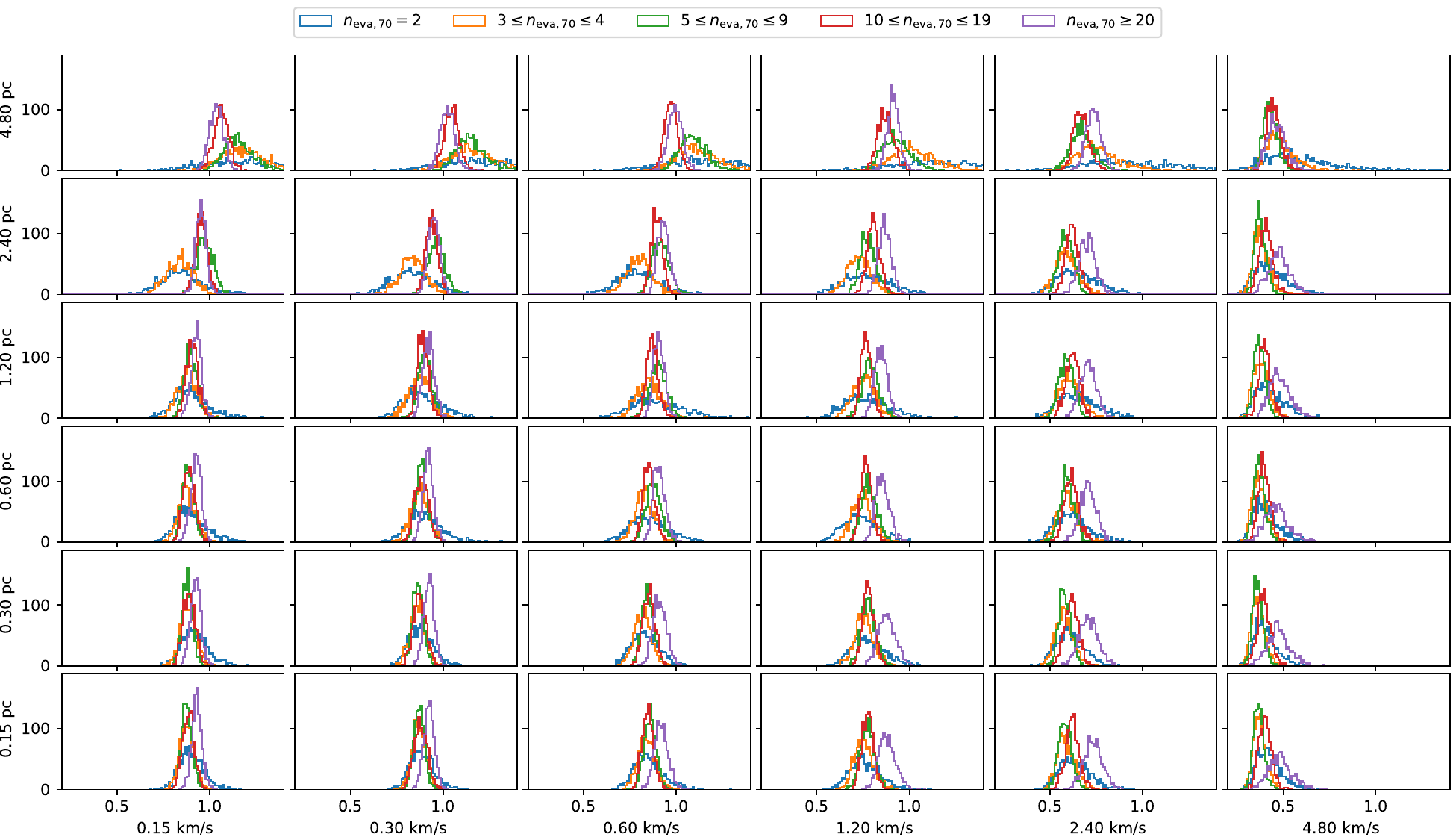}
    \caption{Histograms of $\phi_{\rm eva}$, shown in five bins of $n_{\rm eva,70}$. Only realizations that had at least two clusters in the bin are included.}
    \label{fig:err_median_hist}
\end{figure*}

\begin{figure*}
	\includegraphics[width=2.0\columnwidth]{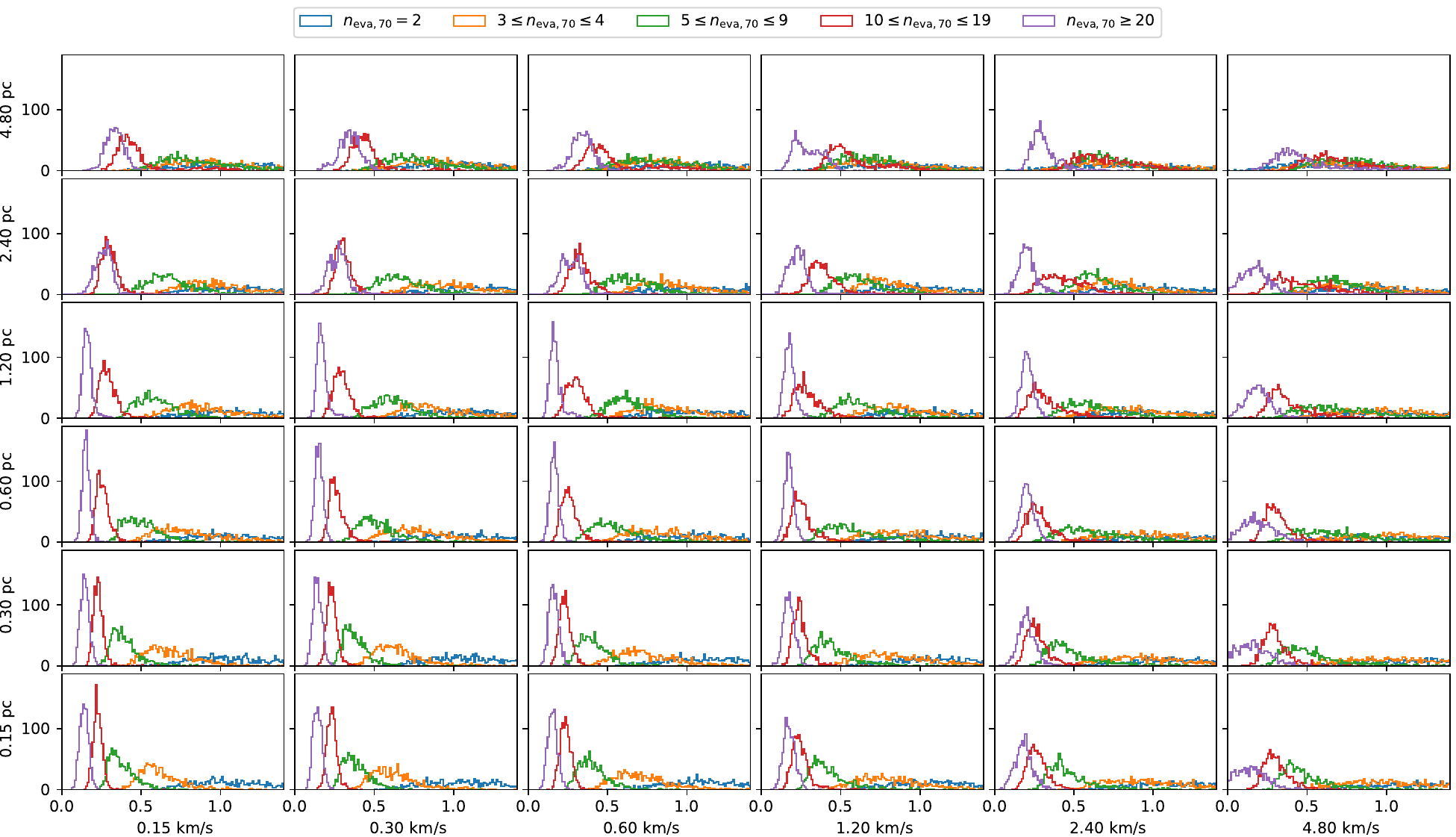}
    \caption{Histograms of $\sigma_{\rm eva}$, shown in five bins of $n_{\rm eva,70}$. Only realizations that had at least two clusters in the bin are included.}
    \label{fig:err_error_hist}
\end{figure*}

As explained in \S\,\ref{w_errors}, we performed 1000 Monte-Carlo realizations of our simulation data for each error pair of position and velocity, $\sigma_{p}$ and $\sigma_{v}$, in a six-by-six grid, assuming a Gaussian distribution of errors. Figures\,\ref{fig:err_median_hist} and \ref{fig:err_error_hist} show $\phi_{\rm eva}$ and $\sigma_{\rm eva}$ in five bins of $n_{\rm eva,70}$. It is clear that for smaller $n_{\rm eva,70}$, the histograms in both cases are broader than for larger $n_{\rm eva,70}$. Also, while $\phi_{\rm eva}$ peaks roughly at the same position, with the exception of $n_{\rm eva,70} \geq 20$ which tends to be slightly shifted toward 1, there is a clear trend in $\sigma_{\rm eva}$ becoming smaller with larger $n_{\rm eva,70}$.

\begin{figure}
	\includegraphics[width=\columnwidth]{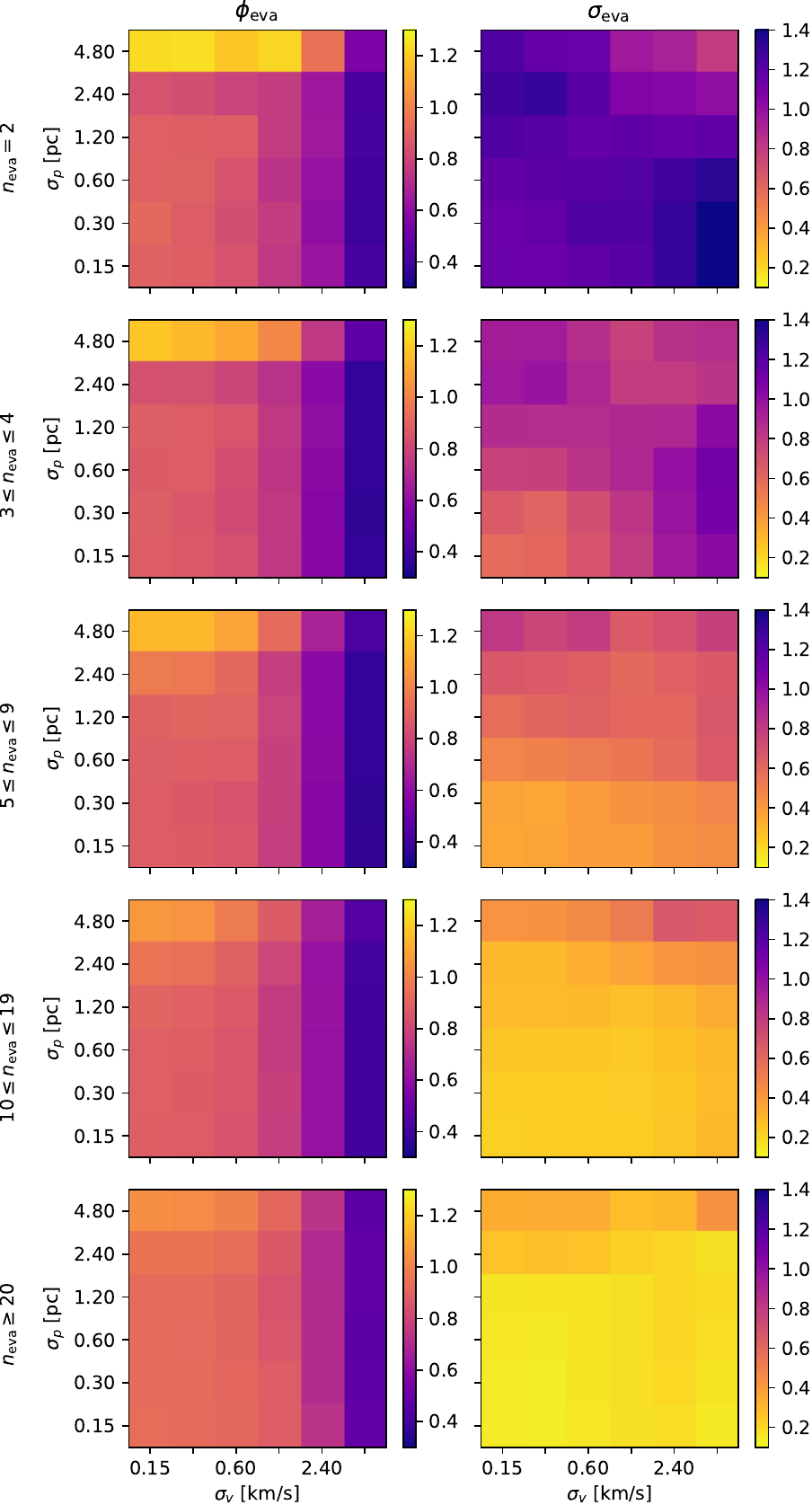}
    \caption{Maps of the medians of $\phi_{\rm eva}$ (left column) and $\sigma_{\rm eva}$ (right column) histograms in the previous figures as functions of $\sigma_v$ and $\sigma_p$, for five bins of $n_{\rm eva,70}$ (rows).}
    \label{fig:error_matrix}
\end{figure}

Figure\,\ref{fig:error_matrix} shows the medians of the previous $\phi_{\rm eva}$ and $\sigma_{\rm eva}$ histograms in 2D maps as functions of $\sigma_{p}$ and $\sigma_{v}$, for each five $n_{\rm eva,70}$ bins. The same information is provided in Table\,\ref{tab:TableB1}\footnote{Table B.1 is only available in electronic form at the CDS via anonymous ftp to cdsarc.u-strasbg.fr (130.79.128.5) or via http://cdsweb.u-strasbg.fr/cgi-bin/qcat?J/A+A/vol/page.}. The values of $\phi_{\rm eva}$ and $\sigma_{\rm eva}$ of the observed clusters are interpolated from these ten maps, using the observed $\sigma_{p}$ and $\sigma_{v}$ in linear space, and $n_{\rm eva,70}$ to select the correct bin (see Table\,\ref{tab:Table1}).

\begin{table}
    \begin{center}
    \caption{Values of the medians of $\phi_{\rm eva}$ and $\sigma_{\rm eva}$ for each error pair in the 1000 Monte-Carlo realizations, in five bins of $n_{\rm eva,70}$. Columns: (1) the lower limit of the $n_{\rm eva,70}$ bin, (2-3) the total observational error in position and velocity, (4) the scaling factor $\phi_{\rm eva}$, (5) the statistical uncertainty $\sigma_{\rm eva}$. Here, we give only a small portion of the table, whose full version is available for electronic download.}
    \label{tab:TableB1}
    \centering
    \setlength{\tabcolsep}{4pt}
    \begin{tabular}{c c c c c }
    \hline \hline
        $n_{\rm eva,70}$ & $\sigma_p$ & $\sigma_v$ & $\phi_{\rm eva}$ & $\sigma_{\rm eva}$ \\
         & [pc] & [km/s] & & \\
         \hline
 2 & 0.15 & 0.15 & 0.896 & 1.140 \\
 2 & 0.15 & 0.30 & 0.883 & 1.134 \\
 \vdots & \vdots & \vdots & \vdots & \vdots \\
  3 & 0.15 & 0.15 & 0.874 & 0.592 \\
 \vdots & \vdots & \vdots & \vdots & \vdots \\
20 & 4.80 & 2.40 & 0.734 & 0.294 \\
20 & 4.80 & 4.80 & 0.461 & 0.431 \\ 
                  \hline
    \end{tabular}\\
    \end{center}
\end{table}

\end{appendix}

\end{document}